\documentclass[twocolumn,preprintnumbers,amsmath,amssymb,
nofootinbib,longbibliography, prb]{revtex4-2}

\usepackage{graphicx}
\usepackage{bm}
\usepackage{dsfont}
\usepackage[usenames,dvipsnames]{xcolor}
\usepackage{pstricks}
\usepackage[tight]{subfigure}
\usepackage{verbatim}
\usepackage{units}
\usepackage{multirow}
\usepackage{mathrsfs}
\usepackage{leftidx}
\usepackage{xspace}
\usepackage{diffcoeff}  
\usepackage{bbm}
\usepackage{amsfonts,amssymb,amsmath}
\usepackage{mathtools}

\usepackage{array} 
\usepackage{tabularx}
\usepackage[normalem]{ulem}
\usepackage{bbold}
\usepackage{pifont}
\usepackage{hyperref}
\hypersetup{colorlinks=true,linktoc=all,linkcolor=blue,breaklinks=true,citecolor=blue,urlcolor=blue}

\usepackage[utf8]{inputenc}
\usepackage[english]{babel}

\definecolor{darkgreen}{RGB}{6, 153, 38}

\addto\captionsenglish{}

\AtBeginDocument{%
    \newwrite\bibnotes
    \def\bibnotesext{Notes.bib}
    \immediate\openout\bibnotes=\jobname\bibnotesext
    \immediate\write\bibnotes{@CONTROL{REVTEX42Control}}
    \immediate\write\bibnotes{@CONTROL{%
    apsrev42Control,author="08",editor="1",pages="1",title="0",year="1"}}
     \if@filesw
     \immediate\write\@auxout{\string\citation{apsrev42Control}}%
    \fi
}%

\newcommand{\beq}{\begin{equation}}
\newcommand{\eeq}{\end{equation}}

\DeclarePairedDelimiterX\braket[2]{\langle}{\rangle}{#1\,\delimsize\vert\,\mathopen{}#2}
\DeclarePairedDelimiterX\ketbra[2]{\lvert}{\rvert}{#1\,\delimsize\rangle\mathopen{}\delimsize\langle\,\mathopen{}#2}

\DeclarePairedDelimiterX\Braket[2]{(}{)}{#1\,\delimsize\vert\,\mathopen{}#2}
\DeclarePairedDelimiterX\Ketbra[2]{\lvert}{\rvert}{#1\,\delimsize)\mathopen{}\delimsize(\,\mathopen{}#2}

\newcommand{\kB}{k_\text{B}}
\newcommand{\kBT}{k_\text{B}T}

\begin{document}

\title{
\vspace*{-1.25cm}
\textnormal{{\small PHYSICAL REVIEW B {\bf 109}, 045429 (2024)}}\\
\vspace*{-0.2cm}
\rule[0.1cm]{18cm}{0.02cm}\\
\vspace*{0.285cm}
Coherent control of thermoelectric currents and noise in quantum thermocouples
}
\author{Jos\'e Balduque}
\affiliation{Departamento de F\'isica Te\'orica de la Materia Condensada, Condensed Matter Physics Center (IFIMAC), and Instituto Nicol\'as Cabrera, Universidad Aut\'onoma de Madrid, 28049 Madrid, Spain\looseness=-1}
\author{Rafael S\'anchez}
\affiliation{Departamento de F\'isica Te\'orica de la Materia Condensada, Condensed Matter Physics Center (IFIMAC), and Instituto Nicol\'as Cabrera, Universidad Aut\'onoma de Madrid, 28049 Madrid, Spain\looseness=-1}
\date{\today}

\begin{abstract}
Three-terminal coherent conductors are able to perform as quantum thermocouples when the heat absorbed from one terminal is transformed into useful power in the other two. Allowing for a phase coherent coupling to the heat source, we introduce a way to control and improve the thermoelectric response via quantum interference. A simple setup composed of a scanning probe between two resonant tunneling regions is proposed that achieves better performance than incoherent analogs by enhancing the generated power and efficiency, and reducing the output current noise.   
\end{abstract}

\maketitle

\section{Introduction}
\label{sec:intro}
The importance of phase coherence in transport through mesoscopic electrical conductors is well known since decades~\cite{Datta1995,nazarov_quantum_2009}. When the system size is small with respect to the length at which carriers decohere in their propagation (i.e., the coherence length), the phases accumulated within the sample induce quantum interference effects that manifest in the mean currents through the system, see e.g., Ref.~\cite{ihn_semiconductor_2009} for a discussion of experimental realizations. The resulting sharp spectral features (e.g. Fabry-P\'erot-like or Fano-like resonances and destructive Aharonov-Bohm interference) can be used as energy filters, one of the requirements to find thermoelectric devices able to efficiently convert heat into electrical power~\cite{hicks_effect_1993,hicks:1993,mahan:1996,whitney_most_2014,sothmann:2015,benenti:2017}. Recent experiments using quantum dots (QDs) have confirmed these expectations~\cite{josefsson_quantum_2018,jaliel:2019}. Phase coherence can then induce a thermoelectric response in conductors that would not manifest it in the absence of interference. Furthermore, this response can be controlled externally in different configurations including magnetic fields~\cite{hofer:2015,samuelsson:2017,haack:2019,haack_nonlinear_2021}, gate voltages~\cite{Vannucci2015}, and movable junctions~\cite{extrinsic}. Interferences have also been used to enhance the thermoelectric efficiency~\cite{finch:2009,karlstrom:2011,gomezsilva:2012,hershfield:2013}.

Multiterminal configurations~\cite{butcher:1990,mazza:2014} contain the nonlocal thermoelectric effect: a charge current is measured in an isothermal conductor formed by two terminals at the same temperature and electrochemical potential out of the conversion of heat injected from one or more other terminals. They introduce the possibility to inject heat directly into the mesoscopic region, allowing one to probe the energetics of internal processes, including electron-electron interactions~\cite{hotspots,thierschmann:2015,dare:2017,walldorf:2017}, phonon-~\cite{entin:2010,jiang:2012,bosisio_nanowire_2016,zhou_three_2020,dorsch:2020,dorsch_characterization_2021} and photon-assisted tunneling~\cite{bergenfeldt:2014}, fluctuating potentials~\cite{sothmann:2012,hartmann:2015,roche:2015}, thermalization~\cite{humphrey_reversible_2005,sanchez:2011,jordan:2013,sothmann:2013,jaliel:2019}, local hotspots~\cite{genevieveqpc,extrinsic} and Kondo correlations~\cite{donsa:2014} in semiconductor devices (with connections with models of hot carrier solar cells~\cite{ross_efficiency_1982,wurfel_solar_1997,wurfel_particle_2005,tesser_thermodynamic_2023}, nothing avoids  the sun to be the third terminal of a mesoscopic conductor), Cooper-pair splitting~\cite{machon_nonlocal_2013,Cao2015,sanchez_cooling_2018,Hussein2019,Kirsanov2019,Tan2021} or Andreev reflection~\cite{mazza:2015,tabatabaei_nonlocal_2022,lopez_optimal_2023} in QDs proximitized with a superconductor, and chirality~\cite{granger_observation_2009,nam_thermoelectric_2013,sanchez:2015qhe} and helicity~\cite{Blasi2020a,blasi_nonlocal_2020b} or nonthermal states~\cite{copro} in quantum Hall edge channels. In most of the above works, the phase coherence of the particles injected from the hot source is lost as they enter the conductor. However, coherently coupling to the hot terminal~\cite{genevieveqpc} makes the source responsible for the thermoelectric response not only for providing heat but also for inducing the necessary energy filtering via quantum interference. Unfortunately, though this effect enhances the longitudinal thermoelectric response substantially~\cite{genevieveqpc}, the nonlocal thermoelectric efficiency remains tiny~\cite{extrinsic}.

\begin{figure}[b]
\includegraphics[width=\linewidth]{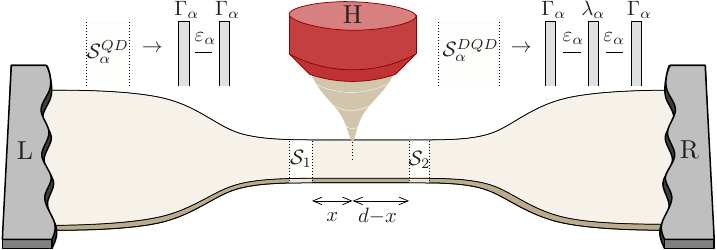}
\caption{\label{fig:scheme}
Scheme of the quantum thermocouple. A quasi one-dimensional quantum conductor connected to two isothermal terminals, L and R, is coupled to a third hot terminal, H, via the tip of a scanning probe. The tip scans the region between two scatterers, represented by the scattering matrices ${\cal S}_\alpha$, $\alpha$=1,2, and separated by a distance $d$. We chose these scatterers to be either resonant tunneling single or double quantum dots with couplings $\Gamma_\alpha$ and $\lambda_\alpha$, and resonance energies $\varepsilon_\alpha$, as indicated in the inset.  
}
\end{figure}

This effect evidences that the coupling to the heat source can dramatically change the conduction properties of a system. Think for instance of using the resulting interference to enhance transport between some pairs of terminals while suppressing it into some others.   
We are interested in finding the simplest configurations where such effect leads to efficient heat to power conversion. For this, we consider a one-dimensional two-terminal conductor with two regions where electrons are scattered (defined by scattering matrices ${\cal S}_\alpha$, with $\alpha$=1,2). The two regions are separated by a distance $d$ short with respect to the electron mean free path (we assume ballistic propagation). This separation can be scanned by the tip of a tunneling microscope~\cite{buttiker:1989,gramespacher:1997,gramespacher:1999} which exchanges particles with the conductor (though we will assume that the injected particle current is zero in average). The tip couples the conductor to a third terminal, H, that acts as the heat source, see Fig.~\ref{fig:scheme}. Transport between every two terminals depends not only on the properties of the two scattering regions $\alpha$ but also on the internal reflections involving the tip. Therefore, the position of the tip, situated at a distance $x$ from the scattering region 1, can be used to shape the transmission probabilities between all three terminals in a way that can be controlled experimentally. Using this property, the system thermoelectric response can be tuned.  Nanometer-resolved nonlocal thermoelectric responses have been achieved in one-~\cite{fast:2020,gachter_spatially_2020} and two-dimensional samples~\cite{Park2013,harzheim:2018,brun2019} via different kinds of scanning probes. 

Here we will consider two cases depending on the chosen structure of the two conductor scattering regions: when they are given by single resonant tunneling QDs, demonstrated experimentally to be efficient energy harvesters~\cite{jaliel:2019}, or by double quantum dots (DQDs). The performance will be quantified in terms of the generated electrical power, the thermoelectric efficiency and the signal to noise ratio of the current output. Ideally, one aims for a conductor that is able to generate a large power output with low noise and high efficiency. In practice, one needs to find a compromise: typically high efficiencies occur at the expense of low power outputs~\cite{benenti:2017}, for instance in the case of very narrow resonances. A practical alternative is to try to improve the efficiency at a given power output that one needs to run a particular device. In that case, systems with boxcar-like transmission probabilities are found to be optimal in two terminal setups~\cite{whitney_most_2014,yamamoto_thermodynamics_2015,bevilacqua_tutorial_2022}. The problem has also been explored in three-terminal coherent conductors, finding similar boxcar transmissions between the conductor terminals and the heat source combined with the extra requirement of having no elastic transfer of particles between the conductor terminals, L and R~\cite{whitney_quantum_2016}. This means that all electrons flowing along the device are forced to visit the heat source connected to the mesoscopic region in their way.
Such broad band transmission probabilities are difficult to find without requiring complicated structures formed by multiple QDs ~\cite{whitney_most_2014,ehrlich_broadband_2021,behera_quantum_2023}. Another possibility is to consider configurations with broken time-reversal symmetry, either by a magnetic field~\cite{sanchez:2011,benenti_thermodynamic_2011,entin_three_2012,brandner_multi_2013,yamamoto_efficiency_2016} or by ac driving~\cite{zhou_boosting_2015,ryu_beating_2022}. A clear interest appears in finding simpler and easier to control configurations.

The current noise is also important in mesoscopic conductors~\cite{blanter_shot_2000,kobayashi_shot_2021}. From an operational point of view, trying to reduce the fluctuations adds a different point of view for optimization~\cite{sanchez:2013,crepieux_mixed_2014,crepieux_heat_2016,ptaszynski_coherence_2018,kheradsoud:2019,timpanaro_hyperaccurate_2023,erdman_pareto_2023}. One can naively expect that boxcar transmissions are also beneficial for reducing the noise signal: the partition noise contribution in two-terminal devices will vanish as transmission is either 0 or 1. Fluctuations in three-terminal configurations are somewhat more complicated and need to be explored.
Note however that, apart from this practical issue, the fluctuations contain fundamental information about the quantum heat engine dynamics~\cite{eymeoud_mixed_2016,eriksson_general_2021,saryal_universal_2021,tesser_charge_2023}. 

We will compare the fully coherent configuration with existing (inelastic) energy harvesters in which carriers are fully thermalized at the heat source~\cite{jordan:2013,jaliel:2019}. They consist of a hot central reservoir directly coupled to the two conductor terminals via (single) resonant tunneling QDs (as in the QD case) achieving efficient energy harvesting. We will refer to them as QD-th in the following.   
Our results show that resonant filtering combined with internal reflections lead to boxcar-shaped transmissions that improve the power, efficiency and signal-to-noise ratio as compared to incoherent case.

The remainder of the paper is organized in the following way. In Sec.~\ref{sec:scattering} we introduce the theoretical description of transport (currents and noises) based on non-interacting scattering theory, as well as the performance quantifiers, and show the resulting zero bias transport properties in Sec.~\ref{sec:currents}. In Sec.~\ref{sec:optim} we discuss the optimization procedure and the results obtained for the different configurations. Conclusions are presented in Sec.~\ref{sec:conclusions}. 

\section{Scattering approach}
\label{sec:scattering}

\subsection{Transport and noise formulas}
\label{sec:trans}

We are interested in the simplest description of transport through multiterminal quantum coherent conductors where correlations due to electron-electron interactions can be disregarded (for related configurations including electron-electron interactions, see e.g., Refs.~\cite{pugnetti_electron_2009,crepieux_electron_2003}). This regime is well described by the Landauer-B\"uttiker scattering theory~\cite{Datta1995,moskalets-book}. Assuming that the system has a single channel, the transport properties between terminals $l$ and $l'$ are encoded in the scattering matrix ${\cal S}_{l'l}$. With it, we can calculate the mean particle and heat currents~\cite{sivan:1986,guttman_thermopower_1995} flowing out of terminal $l$=L,R,H:
\begin{align}
\label{eq:Ipart}
I_l&=\frac{2}{h}\sum_{l'}\int{dE} {\cal T}_{l'l}(E)[f_l(E)-f_{l'}(E)]\\
\label{eq:Jheat}
J_l&=\frac{2}{h}\sum_{l'}\int{dE} (E-\mu_l){\cal T}_{l'l}(E)[f_l(E)-f_{l'}(E)],
\end{align}
where ${\cal T}_{l'l}(E)=|{\cal S}_{l'l}(E)|^2$ is the transmission probability from $l$ to $l'$ and $f_l(E)=\{1+\exp[(E{-}\mu_l)/\kBT_l]\}^{-1}$, the Fermi distribution function of terminal $l$ having an electrochemical potential $\mu_l$ and a temperature $T_l$; the factor 2 accounts for spin degeneracy in the absence of a magnetic field, in which case  we also have that ${\cal T}_{l'l}(E)={\cal T}_{ll'}(E)$. For the charge current, one multiplies Eq.~\eqref{eq:Ipart} by $-e$. Note that the thermoelectric effect requires that the transmission probabilities depend on energy: if all terminals are grounded, $\mu_l=\mu$, $\forall l$, no filtering leads to $I_l=0$, as  $\int{dE}[f_{l}(E)-f_{l'}(E)]=0$. We emphasize that all heat currents are carried by electrons: at low (tens of mK) temperatures, the electron-phonon coupling can be neglected.

We can also calculate the current-current correlations $S_{X_lX_{l'}}(t-t')=\langle\{\Delta \hat X_l(t),\Delta \hat X_{l'}(t')\}\rangle/2$, with $\Delta\hat X_l(t)\equiv\hat X_l(t)-X_l$, where the mean current $X_l=I_l,J_l$ is the statistical average of the current operator: $X_l=\langle\hat X_l(t)\rangle$~\cite{buttiker:1992}. Following Ref.~\cite{blanter_shot_2000} one obtains the zero frequency noise (autocorrelations), which can be decomposed in thermal (th) and shot (sh) noise contributions, $S_{I_lI_l}=S_{I_lI_l}^{\rm th}+S_{I_lI_l}^{\rm sh}$. For a three-terminal configuration ($l,l',m$=L,R,H) they read
\begin{align}
\label{eq:SIIth}
S_{I_lI_l}^{\rm th}=\frac{4}{h} \int dE \sum_{l'\neq l} \mathcal{T}_{ll'}[f_l(1-f_{l})+f_{l'}(1-f_{l'})]
\end{align}
and
\begin{gather}
\label{eq:SIIsh}
\begin{aligned}
S_{I_lI_l}^{\rm sh} = \frac{4}{h} \int dE &\left[\sum_{{l'},m\neq{l}}\frac{{\cal T}_{ll'}{\cal T}_{lm}}{2}(f_{l'}-f_m)^2\right.\\
&+\left.\sum_{l'}{\cal T}_{ll}{\cal T}_{l'l}(f_{l'}-f_l)^2\right].
\end{aligned}
\end{gather}
For the heat noise, $S_{J_lJ_l}$~\cite{sergi_energy_2011,zhan_electronic_2011}, one needs to include $(E-\mu_l)^2$ in the integrals of the above Eqs. \eqref{eq:SIIth} and \eqref{eq:SIIsh}. The expression for heat-charge crosscorrelations (aka mixed noise) is slightly more cumbersome:
\begin{align}
\label{eq:SIJ}
S_{I_l J_{l'}}= & \frac{4}{h}\int dE (E{-}\mu_{l'}) \Big\{ \sum_{r}{\cal T}_{l r}{\cal T}_{l' r}f_r(1-f_r)\nonumber\\
&  - {\cal T}_{l l'} [f_l(1-f_l)+f_{l'}(1-f_{l'})]\\
& + \frac{1}{4} \sum_{r\neq r'}{\cal M}_{ll',rr'} [f_r(1{-}f_{r'})+f_{r'}(1{-}f_r)]\Big\},\nonumber
\end{align}
where we have defined
\beq
{\cal M}_{ll',rr'}\equiv{\cal S}^{\dagger}_{lr}{\cal S}_{lr'}^{}{\cal S}^{\dagger}_{l'r'}{\cal S}_{l'r}^{}+{\cal S}_{lr}^{}{\cal S}_{lr'}^{\dagger}{\cal S}_{l'r'}^{}{\cal S}_{l'r}^{\dagger}.
\eeq
In the following we will concentrate on correlations including the heat current injected from the tip and the generated current in terminal R. Hence, we will simplify the notation as $S_{II}\equiv S_{I_RI_R}$, $S_{JJ}\equiv S_{J_HJ_H}$ and $S_{IJ}\equiv S_{I_RJ_H}$.

\subsection{Scattering components}
\label{sec:performance}

We will focus on a three-terminal conductor working as an energy harvester as the one depicted in Fig.~\ref{fig:scheme}. A one-dimensional elastic conductor composed of two terminals, $l$=L,R, at the same temperature, $T$, and with electrochemical potential $\mu_l$ are connected by two scattering regions $\alpha$=1,2 with scattering matrices
\begin{equation}
\label{eq: S_scatterers}
\mathcal{S}_{\alpha}=   
\begin{pmatrix}
    r_{\alpha} & \tau_{\alpha} \\
     \tau_{\alpha} & r_{\alpha}
\end{pmatrix},
\end{equation}
where $r_\alpha$ and $\tau_\alpha$ are the reflection and transmission amplitudes, respectively, and fulfill $|r_\alpha|^2+|\tau_\alpha|^2=1$. We consider single-channel junctions, for simplicity. We will investigate two different configurations (sketched in Fig.~\ref{fig:scheme}) where the two scattering regions are either a resonant tunneling QD, with Breit-Wigner amplitudes~\cite{buttiker:1988},
\begin{gather}
\begin{aligned}
\label{eq:tau_qd}
    \tau_\alpha^{QD}(E) &= \frac{-i\Gamma_{\alpha}}{E-\varepsilon_\alpha + i\Gamma_\alpha}\\
    r_\alpha^{QD}(E) &= 1+\tau_\alpha^{QD}(E),
\end{aligned}
\end{gather}
with resonance energy $\varepsilon_\alpha$ and broadening $\Gamma_\alpha$, or a DQD~\cite{sumetskii_absolute_1991} (see Appendix~\ref{sec:dqdscatt} for details),
\begin{gather}
\begin{aligned}
\label{eq:tau_dqd}
    \tau_\alpha^{DQD}(E) &= -\frac{i\Gamma_{\alpha} \lambda_\alpha}{(E-\varepsilon_\alpha + i\Gamma_\alpha/2)^2-\lambda_\alpha^2}\\
    r_\alpha^{DQD}(E) &=1 + \frac{(E-\varepsilon_\alpha + i\Gamma_\alpha/2)}{\lambda_\alpha}\tau_\alpha^{DQD}(E),
\end{aligned}
\end{gather}
where $\lambda_\alpha$ is the interdot coupling. We assume symmetric DQDs, each one described by a single resonance energy, $\varepsilon_\alpha$, and coupling $\Gamma_\alpha$, for simplicity. We will later also consider that $\Gamma_1=\Gamma_2=\Gamma$ (both in the QD and DQD cases) and $\lambda_1=\lambda_2=\lambda$. Noninteracting DQDs have recently been used to indicate deviations of the fluctuations in coherent transport from classically established thermodynamic bounds~\cite{agarwalla_assessing_2018,prech_entanglement_2023}. 

The two regions are separated by a distance $d$. A third terminal, H, at temperature $T_H=T+\Delta T$ and electrochemical potential $\mu_H$ is coupled to the conductor via the tip of a scanning tunneling probe at a distance $x<d$ of the scattering region 1. Scattering at the conductor-tip junction is given by the matrix~\cite{buttiker:1984,buttiker:1989} 
\begin{equation}
\label{eq: S_tip}
        \mathcal{S}_{\rm tip}=   \begin{pmatrix}
        -\eta_{-}/2 & \sqrt{\epsilon} & \eta_{+}/2 \\
        \sqrt{\epsilon} & \eta_{-} - 1 & \sqrt{\epsilon} \\
        \eta_{+}/2 & \sqrt{\epsilon} & -\eta_{-}/2
        \end{pmatrix},
\end{equation}
where $\eta_{\pm}=1 \pm \sqrt{1-2\epsilon}$. The real parameter $\epsilon \in[0,1/2]$ quantifies the tip-conductor coupling. The tip-conductor geometry permits to uncouple the heat source when $\epsilon=0$, in which case the system behaves as an isothermal two-terminal conductor. 

The hot terminal is assumed to be a voltage probe~\cite{buttiker_four_1986} with a floating electrochemical potential which will adapt to the condition $I_H=0$. This way, heat but no net charge is injected through the tip into the conductor. Note that the heat injection mechanism is fully coherent and adds to the interferences due to internal reflection in the conductor. Electrons propagating within scattering regions with wavenumber $k(E)=\sqrt{2m(E-U_0)}/\hbar$ will accumulate a kinetic phase $\chi_x(E)/2=k(E)x$ in their way between 1 and the tip, and $\chi_{d-x}(E)/2=k(E)(d-x)$ between the tip and 2. We take the energy of the lowest subband of the one-dimensional conductor~\cite{ihn_semiconductor_2009} as the energy origin, $U_0=0$. In the ballistic regime, we can safely neglect disorder effects in the nanowire potential\footnote{See Refs.~\cite{bosisio_using_2015,muttalib_nonlinear_2015} for related works discussing diffusive thermoelectricity in disordered nanowires.}.
The region between ${\cal S}_1$ and ${\cal S}_2$ can be viewed as a double Fabry-P\'erot interferometer with a movable intermediate barrier which has the particularity that it absorbs and reinjects particles.
The energy dependence of the accumulated phases are sufficient to induce a thermoelectric effect in interferometers~\cite{hofer:2015,Vannucci2015,samuelsson:2017,haack:2019,haack_nonlinear_2021,extrinsic} or to enhance it~\cite{finch:2009,gomezsilva:2012,hershfield:2013}.   

\subsection{Shaping the transmission probabilities}
\label{sec:compose}

Known the scattering matrices of the two scattering regions and the tip junction, Eqs.~\eqref{eq: S_scatterers} and \eqref{eq: S_tip}, as well as the phases accumulated in the connections between them, $\chi_{d-x}$ and $\chi_x$, we are ready to obtain the scattering matrix of the whole system. For this one simply needs to take into account that e.g., the right-outgoing wave from region 1 is the left-ingoing wave at the tip, multiplied by a phase factor ${\rm e}^{i\chi_{x}(E)/2}$, and the same for the wave going out of the tip and scattering at 2 with a factor ${\rm e}^{i\chi_{d{-}x}(E)/2}$ instead, and solve for all the outgoing waves as functions of the ingoing ones, see e.g., Ref.~\cite{Datta1995}, or Ref.~\cite{extrinsic} for a more related configuration.

The resulting scattering matrix ${\cal S}(E)$ connects the three terminals L, R and H. The obtained transmission probabilities can be decomposed as:
\begin{equation}
\label{eq: transmissionslh}
    \mathcal{T}_{lH}(E) =2\epsilon|\mathcal{A}|^{-2}|{\tau}_{\alpha_l}|^2\mathcal{I}_{lH}, 
\end{equation}
for trajectories between terminals $l$=L,R and H, with $\alpha_{L(R)}$=1(2), and
\begin{equation}
\label{eq: transmissionslr}
    \mathcal{T}_{LR}(E) = \frac{\eta_{+}^2}{4|\mathcal{A}|^2}|{\tau}_1|^2|{\tau}_2|^2,
\end{equation}
for elastic transport along the conductor (without visiting the probe terminal). In these expressions,  $\tau_\alpha$ are the transmission amplitudes at scatterer $\alpha$, see Eqs.~\eqref{eq:tau_qd} and \eqref{eq:tau_dqd}, while
\begin{equation}
\label{eq: A}
    \mathcal{A}=  1+\frac{\eta_{-}}{2}\left(r_1e^{i\chi_x}{+}r_2e^{i\chi_{d{-}x}}\right) - \sqrt{1{-}2\epsilon} \; r_1 r_2 e^{i\chi_d},
\end{equation}
where $\chi_d=\chi_x+\chi_{d-x}$, and
\begin{equation}
\label{eq: Interference_pattern}
\mathcal{I}_{l H}(E) =  1-\frac{|\tau_{\beta_{l}}|^2}{2} + {\rm Re}[r_{\beta_{l}} e^{i\chi_l}]
\end{equation}
account for the quantum interference of multiple internally reflected trajectories between the QDs and the tip,
with  $\chi_{L(R)}=\chi_{d-x}(\chi_x)$ and $\beta_{L(R)}$=2(1). As the scattering matrix is unitary, reflection probabilities can be calculated from the transmissions: ${\cal T}_{ll}=1-\sum_{l'\neq l}{\cal T}_{ll'}$.

\begin{figure}[t]
\centering
\includegraphics[width=\linewidth]{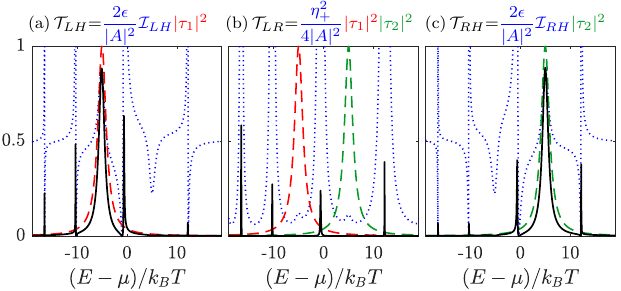}
\caption{\label{fig:trans}  Transmission probabilities between  (a) terminals $L$ and $H$, (b) terminals $L$ and $R$ and (c) terminals $R$ and $H$ as a function of the energy of the scattered electrons when the scattering regions 1 and 2 correspond to the single QD case. The total transmission (black full line) can be decomposed as suggested in Eqs. \eqref{eq: transmissionslh} and \eqref{eq: transmissionslr} (red/green dashed: transmission probability of quantum dots $1,2$; blue dotted: contribution from the multiple internally reflected trajectories between scattering regions and the tip). Parameters: $d=10l_0, \; x=d/2, \; \epsilon=0.5, \; \Delta E = 10\kBT, \; \Gamma = \kBT, \; \mu=20\kBT$.}
\end{figure}

As an illustration, the three transmissions are plotted in Fig.~\ref{fig:trans} for the case of QD scatterers when the tip is right in the middle distance between them. Henceforth, lengths are expressed in units of $l_0=\hbar/\sqrt{8m\kBT}$, which for GaAs wires is of the order of \unit[40]{nm} at $T=\unit[1]{K}$. We will also assume that the tip is strongly coupled, $\epsilon\lesssim0.5$, which gives the largest power~\cite{extrinsic}. While $|\tau_\alpha^{QD}|^2$ give Lorentzian line shapes around the energies $\varepsilon_1$ (red-dashed line) and $\varepsilon_2$ (green-dashed line), the interference patterns (blue-dotted lines) strongly influence the total transmission probabilities (full black lines), which show Fano-shaped narrower tunneling resonances at $\varepsilon_\alpha$ and additional very sharp peaks. To understand the Fano-like interference, one can think of the closed trajectories formed between the tip and the reflecting barrier (scatterer 2 for ${\cal T}_{LH}$, scatterer 1 for ${\cal T}_{RH}$) as playing the role of a localized state in parallel to trajectories that are transmitted through the other barrier and the tip. Of particular interest for our purposes here is the destructive interference that strongly suppresses the elastic contribution between L and R, see Fig.~\ref{fig:trans}(b). 

 Note that the developed interference patterns require scatterers with broad spectral features. In this sense, the weak coupling limit where $\Gamma_\alpha\ll\kBT$, so $|\tau_\alpha^{QD}|^2\approx\Gamma_\alpha\delta(E-\varepsilon_\alpha)$, is not particularly useful.

\subsection{Thermoelectric response}
\label{sec:response}

Transport between the two conductor terminals, L and R, consists of two contributions. On top of the elastic transmission between L and R, ${\cal T}_{LR}(E)$, sequential transmission between the conductor terminals and the tip, given by ${\cal T}_{LH}(E)$ and ${\cal T}_{HR}(E)$, introduces an inelastic contribution. The current can be decomposed in the two contributions, $I_R=I_{\rm el}+I_{\rm inel}$, with 
\begin{align}
I_{\rm el}&=\frac{2}{h}\int{dE} {\cal T}_{LR}(E)[f_R(E)-f_L(E)]\\
I_{\rm inel}&=\frac{2}{h}\int{dE} {\cal T}_{HR}(E)[f_R(E)-f_H(E)],
\end{align}
with particle conservation $I_L=-I_R$ being guaranteed by the probe condition, $I_H=0$.
In the absence of an applied voltage, the elastic contribution vanishes (remember $T_L=T_R$). Even worse, as a finite bias is applied, the elastic contribution will always flow downhill, i.e., dissipating (Joule) power. Hence, the thermoelectric power generation relies on the inelastic contribution~\cite{sothmann:2015,wang_inelastic_2022}. Note that the inelastic scattering contribution is not describing heat leakage with an environmental bath but rather with an electronic reservoir that is treated in equal footing as the conductor terminals.

At zero bias, a finite current will flow as soon as the energy dependence of ${\cal T}_{LH}(E)$ and ${\cal T}_{HR}(E)$ are different. The sign of the current will be determined by the dominance of hole- or electronlike character of the two lead-tip transmissions, similar from what one expects in a conventional thermocouple~\cite{benenti:2017}. In our case, this can be controlled with the energy gain $\Delta E\equiv\varepsilon_2-\varepsilon_1$ given by the position of the resonances (measured with respect to the equilibrium electrochemical potential, $\mu$), and tuned by gate voltages. It is also modulated with the internal interferences depending on the position of the tip, $x$.

Then, a finite power, $P=-I_R\Delta\mu>0$, will be generated as long as the inelastic contribution flowing against the electrochemical potential difference $\Delta\mu\equiv\mu_R-\mu_L$ (applied symmetrically around $\mu$) is larger than the elastic one flowing in favor. The maximal generated power in a quantum channel connecting two electronic reservoirs with different temperatures turns out to be limited by quantum mechanical effects~\cite{pendry_quantum_1983}, as found in Refs.~\cite{whitney_most_2014,whitney_finding_2015}, with the bound given by:
\begin{equation}
    \label{eq: power_bound}
    P_W=2A_0\frac{\pi^2}{h}\kB^2\Delta T^2,
\end{equation}
where $A_0\simeq 0.0321$, the factor 2 accounts for spin degeneracy and $\Delta T$ is the temperature difference between reservoirs. This is a two-terminal result, so to compare with an optimal thermocouple heat engine, that consists of two systems with opposite
thermoelectric responses, we have to multiply by a factor 2.

Another way to quantify the thermoelectric performance is the efficiency, $\eta=P/J_H$, whose thermodynamic bound is, not very surprisingly~\cite{whitney_quantum_2016}, established by the Carnot efficiency, $\eta_C=1-T/T_H$. However, it is not the main limitation in this case, as quantum mechanical effects establish a stronger bound for the efficiency at any given power output, $\eta_W$ \cite{whitney_most_2014}. Although there is not analytical expression for this value, it is possible to obtain it numerically following Ref.~\cite{whitney_finding_2015} by finding the narrower boxcar transmission that results in the chosen power, which also depends on the choice of applied bias. One usually needs to find a compromise between high power output and high efficiency, which may depend on the desired operation at task.  

Additionally, one wants that the generated power does not strongly fluctuate. This question has generated strong interest in the last years related to thermodynamic bounds on the amount of fluctuations depending on the entropy produced by the engine. These are the so-called thermodynamic uncertainty relations, initially obtained for classical stochastic engines~\cite{barato_tur_2015} and shown to be modified in the presence of quantum coherence~\cite{brandner_thermodynamic_2018,agarwalla_assessing_2018,potanina_thermodynamic_2021} and generalized for multiterminal configurations~\cite{dechant_multidimensional_2018,lopez_optimal_2023}. Here we will adopt a more practical strategy and quantify the signal to noise ratio via the particle current inverse Fano factor~\cite{blanter_shot_2000}:
\beq
\frac{1}{F}=\frac{2|I_R|}{S_{I_RI_R}}.
\eeq
Note that while the isothermal transport of noninteracting electrons is sub-Poissonian ($F<1$)~\cite{blanter_shot_2000}, there is no such restriction in the presence of a temperature difference: think for instance of the stopping voltage where current vanishes on average but fluctuations do not, see e.g. Ref.~\cite{kheradsoud:2019}.
We will not compare with any complicated bound; here we are simply interested in trying to keep $1/F$ as high as possible.

\subsection{Thermalized cavity model}
\label{sec_qth}
To emphasize the quantum coherent effects, we will compare the performance of the above models with an energy harvester where the two scatterers (formed by single resonant tunneling QDs) couple the two conductor terminals directly to the hot reservoir. This way, electrons are thermalized in H as they tunnel from the conductor terminals through the QDs. Hence, the elastic contribution vanishes, as ${\cal T}_{LR}^{QD-th}(E)=0$, and currents are fully determined by the transmission probabilities ${\cal T}_{lH}^{QD-th}(E)=\Gamma_{\alpha_l}^2/[(E-\varepsilon_{\alpha_l})^2+\Gamma_{\alpha_l}^2]$ which enter the inelastic transport~\cite{jordan:2013}. Note this model cannot be obtained from the limit of a strongly coupled tip ($\epsilon\rightarrow1/2$) in the QD setup: terminal H requires two channels to avoid interference~\cite{buttiker:1986}.   

\section{Zero bias transport}
\label{sec:currents}

\begin{figure}[t]
    \includegraphics[width=\linewidth]{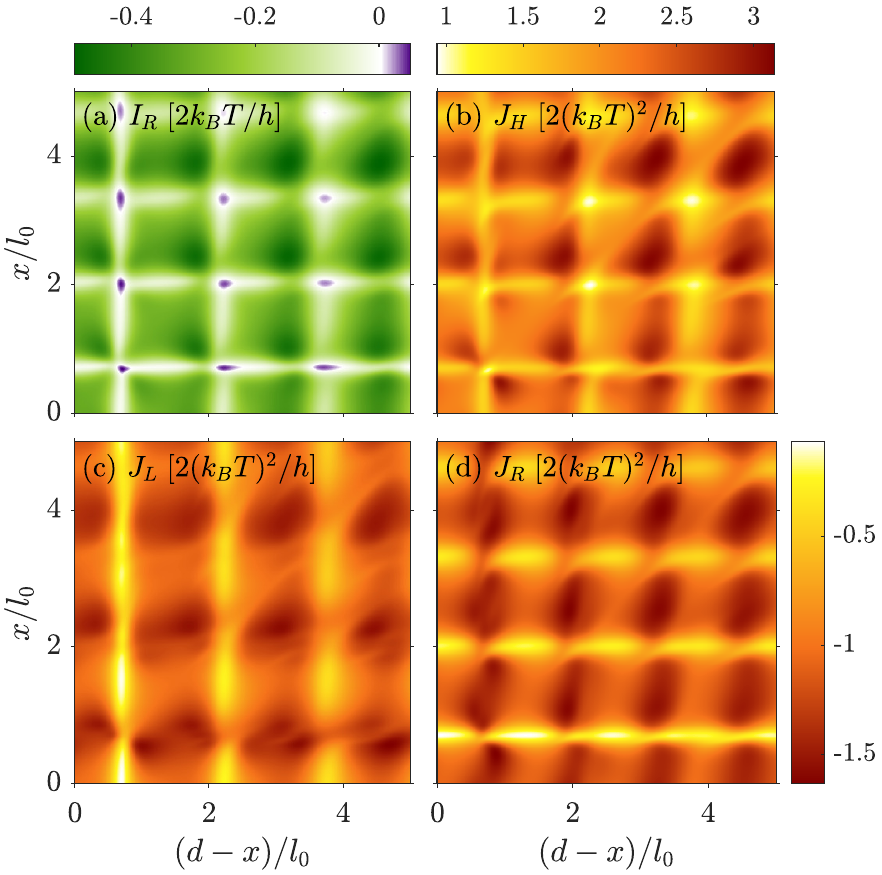}    \caption{\label{fig:currents1} (a) Generated thermoelectric current, $I_R$, and (b)-(d) heat currents in reservoirs $l$=H,L,R, $J_l$, when the temperature of reservoir H is increased by $\Delta T/T = 1$ and the scatterers are single quantum dots, with $\Delta\mu=0$. Parameters: $\Delta E=6\kBT$, $\epsilon=0.5$, $\Gamma = 2\kBT$, $\mu=20\kBT$. }
\end{figure}
Let us first consider the transport properties of the unbiased conductor with $\mu_L=\mu_R=\mu$. Without coupling to the hot reservoir ($\epsilon=0$), all currents are zero, as expected for a system in equilibrium. Only noise will be finite due to the thermal fluctuations, see Eq.~\eqref{eq:SIIth}. As the tip is coupled, electrons in the conductor at a distribution $f_L(E)=f_R(E)$ can be absorbed by terminal H and be reinjected with a hotter distribution $f_H(E)$, leading to a heat current $J_H$. This process breaks detailed balance provided ${\cal T}_{LH}(E)\neq{\cal T}_{RH}(E)$: Consider for instance the configuration shown in Fig.~\ref{fig:trans}, where ${\cal T}_{LH}(E)$ is on average more weighted for energies below the electrochemical potential, with the opposite being true for ${\cal T}_{RH}(E)$, for instance if $\Delta E>0$. Then, cold electrons from L will most likely be absorbed by the tip than those from R; similarly, the reinjected hot electrons will most likely be transferred to R than to L. As a result, a net current will flow from L to R. The position of the tip modulates the energy dependence transmission probabilities ${\cal T}_{lH}(E)$, resulting in periodic oscillations which suppress the generated current. Eventually the current changes sign for low $\Delta E$. This is shown in Fig.~\ref{fig:currents1}(a) as a function of the distances from the tip to the two scatterers. This is an effect of interference, not present in the QD-th configuration with similar QDs~\cite{jordan:2013}. Similar oscillations show up in the heat currents~\cite{extrinsic} flowing out of the hot terminal and into the conductor ones, with the maxima of all four currents occurring in the same regions, see Figs.~\ref{fig:currents1}(b)-(d). Note that in some regions all currents are suppressed, suggesting that the tip works as a valve. Note also that the heat currents are suppressed in the symmetric condition $x=d/2$, with replicated features parallel to this condition. Remarkably, the spots where $I$ changes sign coincide with the minima of the injected $J_H$, which may be attributed to resonances close to $\mu$ that require a smaller (and opposite) energy gain than $\Delta E$ to generate current from R to L.

\begin{figure}[t]
    \includegraphics[width=\linewidth]{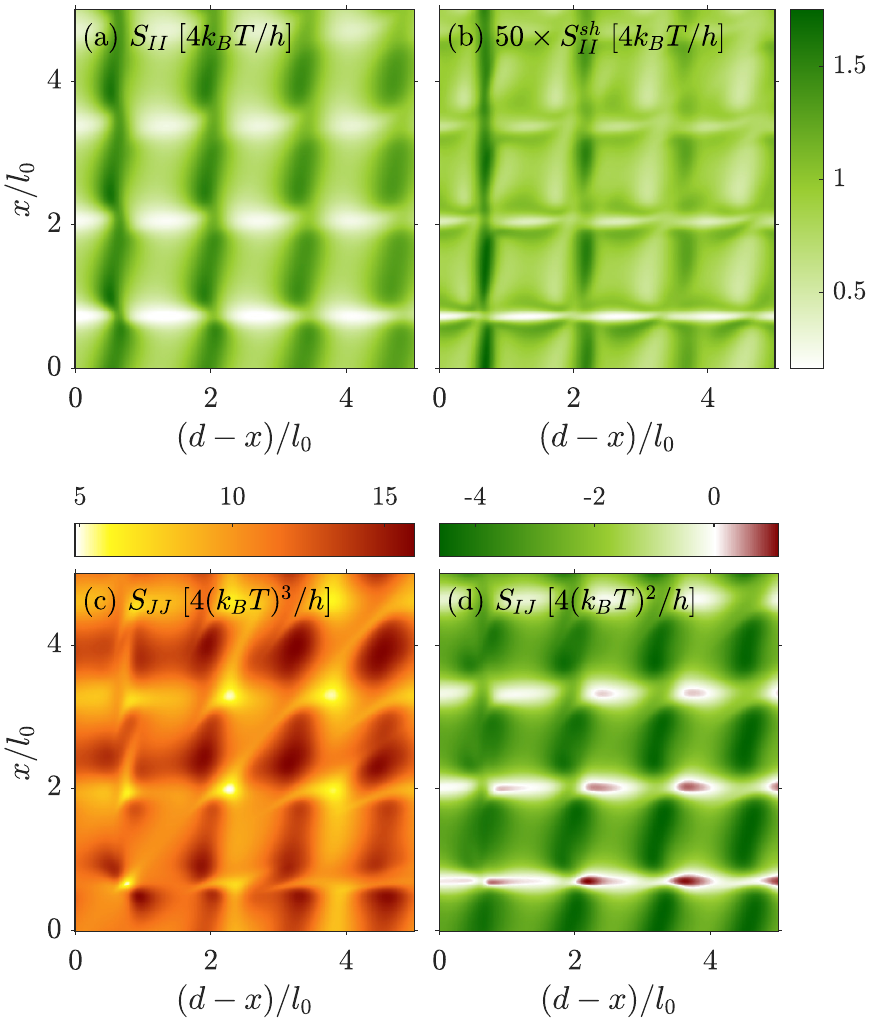}
    \caption{\label{fig:noises1} (a) Particle current noise and (b) its shot contribution at terminal R, (c) heat noise at terminal H, and (d) crosscorrelation between injected heat from H and particle current at R, for the same scenario and parameters of Fig.~\ref{fig:currents1}.} 
\end{figure}
Interestingly, features that look similar in the generated current and the injected heat, like e.g., the current suppression for fixed $x$ and for fixed $d-x$, see vertical and horizontal regions in Figs.~\ref{fig:currents1}(a) and~\ref{fig:currents1}(b), have very different noises, as shown in Fig.~\ref{fig:noises1}. Indeed $S_{II}$ is maximal along the vertical ones (for fixed $d-x$) and minimal along the horizontal (fixed $x$), see Fig.~\ref{fig:noises1}(a). This asymmetry, caused by the finite $\Delta E$, is much less evident in the fluctuations of the injected heat current, $S_{JJ}$, cf. Fig.~\ref{fig:noises1}(c). 
Though $S_{II}$ is strongly dominated by the thermal contribution at these temperatures, the shot noise term is also sensitive to this feature, see Fig.~\ref{fig:noises1}(b). Note that $S_{II}^{sh}$ is rapidly suppressed as the separation of the two scatterers increases. This is not the case of the mixed noise, which increases with the scatterers' separation, thus serving as an indicator of the tiny spots where the generated current changes sign in the long-wire regime, see Fig.~\ref{fig:noises1}(d).
Let us mention that the noise $S_{I_LI_L}$ shows the same features as $S_{I_RI_R}$ reflected over the antidiagonal: i.e., maxima in the horizontal current suppressions, minima in the vertical ones (not shown). However, in the regions where the particle current is maximal, both contributions are roughly the same, so it does not matter which one we chose for the optimization.

\begin{figure}[t]
    \includegraphics[width=\linewidth]{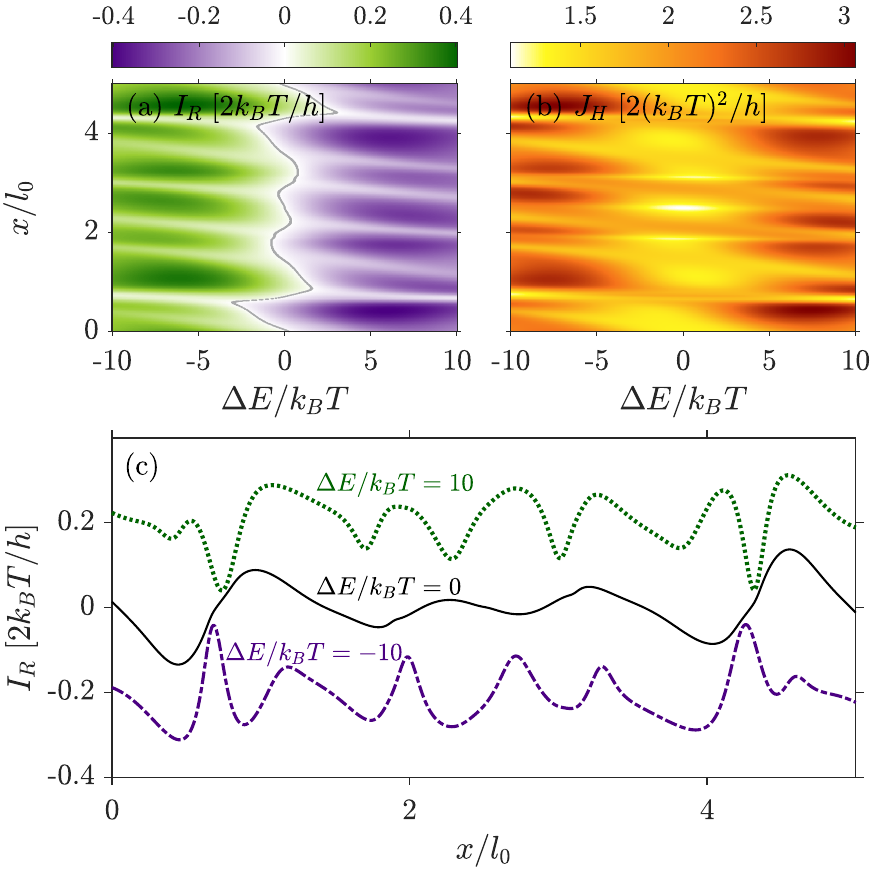}
    \caption{\label{fig:currents} (a) Generated zero bias thermoelectric current, $I_R$ and (b) injected heat current, $J_H$, as functions of $x$ and $\Delta E$ in the QD case. The dashed line in (a) marks the condition $I=0$. (c) Cuts of the current in panel (a) for fixed $\Delta E$, as indicated. Parameters: $d=5l_0$, $\epsilon=0.5$, $\Gamma = 2\kBT$, $\mu=20\kBT$, $\Delta T/T = 1$. }
\end{figure}
In a real wire, the separation of the scatterers is typically fixed, while the position of the resonant energies can be controlled with plunger gates~\cite{jaliel:2019}. Figures~\ref{fig:currents}(a) and \ref{fig:currents}(b) show the generated current and the injected heat as functions of the QD energy difference $\Delta E$ and the position of the tip, $x$, for a fixed distance $d=5l_0$. The sign of the current is mainly determined by the sign of $\Delta E$, as expected, see also Fig.~\ref{fig:currents}(c). However, for small differences, the interference induced by the position of the tip is able to reverse the sign of the current. At $\Delta E=0$, where a thermalized 2DEG thermocouple would be electron-hole symmetric (and hence $I=0$~\cite{jordan:2013}), the scattering at the hot tip is able to induce an oscillating current, as shown explicitly in Fig.~\ref{fig:currents}(c). Note the expected symmetry $X_l(x,\Delta E)=-X_l(d{-}x,-\Delta E)$ is fulfilled both for $I_R$ and $J_H$.

The maxima of the generated current and the injected heat occur at different positions, however in both cases they are in the region around $\Delta E\sim5-10\kBT$, for $T_H=2T$, see Figs.~\ref{fig:currents}(a) and \ref{fig:currents}(b). At higher energies of the resonant levels, the thermal fluctuations of terminal H become negligible and both currents are suppressed. This result also depends strongly on other parameters, in particular, the linewidth, $\Gamma$ (here we considered $\Gamma=2\kBT$ which maximizes the response, as will be shown later). For this reason, it is useful to find the optimal parameters that enhance the thermoelectric properties of the device. In what follows we will do so for the two configurations consisting of QD and DQD scatterers.
We will (quite arbitrarily) consider $\mu=20\kBT$ and $\Delta T/T=1$, except where explicitly stated.

\section{Performance optimization}
\label{sec:optim}

\begin{figure}[t]
    \includegraphics[width=\linewidth]{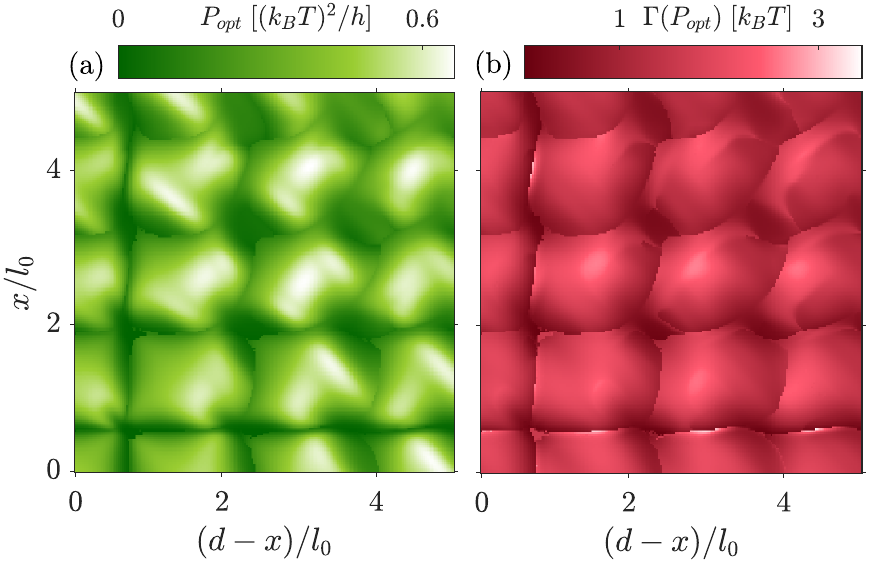}
\caption{\label{fig:optimization} Optimization procedure: the parameters of the QD system, $\Lambda_{QD}$, are optimized for each configuration $(d-x,x)$ to obtain the maximum power, $P_{opt}$, shown in panel (a) for fixed $\Delta T/T=1$ and $\mu=20\kBT $. (b) Obtained values for $\Gamma$.}
\end{figure}

In the sections above, transport was induced by the hot terminal, with the electric current flowing at zero bias. To generate a finite power, this current needs to flow against an electrochemical potential difference. In the following, we focus on the performance of such a heat engine. There are different ways one can optimize it. Traditionally, researchers have been interested in getting either the highest efficiency~\cite{mahan:1996}, which usually comes along with tiny generated powers, or the efficiency at maximum power, which may be a good strategy when one is not limited by the amount of generated power and wants to minimize the cost of the used resources, e.g. in power plants~\cite{curzon:1975}. A different strategy is to maximize the efficiency for a given value of power that one needs to use~\cite{whitney_most_2014}, for instance, to run a particular device with an on-chip energy harvester. Depending on the chosen procedure, the desired properties of the engine are different: sharp resonant transmissions give maximal efficiencies, while boxcar-shaped transmissions maximize the efficiency for a given power.

In a mesoscopic sample as ours, the generated power is expected to be a small quantity. Also, as we discuss below, the different quantities cannot be optimized simultaneously. Hence, we chose to optimize the parameters that give the maximal power and compare the resulting performance quantifiers (power, effficiency and noise) for the different models with QD and DQD scatterers and the QD-th model of Ref.~\cite{jordan:2013}.

Let us first describe the procedure to find optimal configurations of the device in detail. 
For fixed $\mu$ and $\Delta T$, we define the function $P(x,d,\Lambda_q)$ which gives the extracted power for given distances $x$ and $d$ and a particular set of parameters, $\Lambda_q$, we want to optimize: $\Lambda_{QD}=\{\Gamma, \Delta E, \Delta \mu, \epsilon\}$, $\Lambda_{DQD}=\{\Gamma,\lambda, \Delta E, \Delta \mu, \epsilon\}$ and $\Lambda_{QD{-}th}=\{\Gamma, \Delta E, \Delta \mu\}$, in each case. Note that $\Lambda_q$ fully describe the properties of the scatterers. Then, the function $-P(x,d,\Lambda_q)$ is numerically minimized with respect to the chosen set. This is repeated for multiple combinations of distances, as illustrated in Fig. \ref{fig:optimization}(a) for the QD case, where $P_{opt}=P(\Lambda^{opt}_{QD})$ is the value of the extracted power for the parameters obtained after the minimization, $\Lambda_{QD}^{opt}$. As a representative value, we show in Fig. $\ref{fig:optimization}$(b) the obtained values of $\Gamma_{opt}$ that optimize power at each pair of distances $(d-x,x)$, which is found to vary around a few times $\kBT$. Note that for the distances where the currents are suppressed (see Fig.~\ref{fig:currents1}) no set of parameters is able to generate any substantial power. In these regions, the destructive interference between the two scatterers (strongly dependent on $x$ and $d$) dominates (the absence of) transport. Furthermore, for the areas where $P_{opt}$ is maximum, the optimal values for the parameters (here only shown for $\Gamma$) only vary smoothly. 

\begin{figure}[t]
\includegraphics[width=\linewidth]{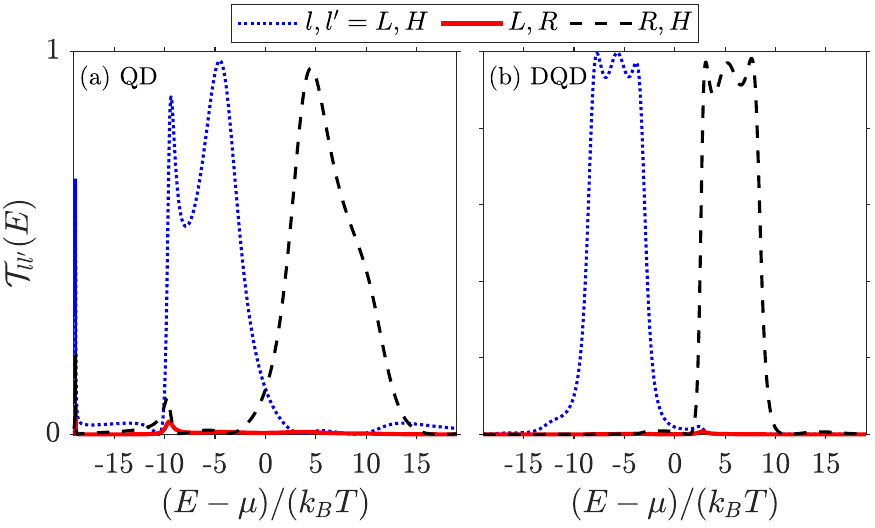}
  \caption{\label{fig:opt_trans} Energy dependence of the (a) QD and (b) DQD transmission probabilities when parameters are chosen to optimize the extracted power at a particular position of the tip having a maximum of $P$: (a) $d{=}5.6l_0$, $x{=}2.6l_0$, giving $\Lambda_{QD}/\kBT=\{2.3,11,1.6,0.5/\kBT\}$, and (b) $d{=}6.4l_0$, $x{=}6.2l_0$, with $\Lambda_{DQD}/\kBT=\{3.1,2.1,11.7,2.2,0.43/\kBT\}$.
    } 
\end{figure}

Our numerical minimization searches for local minima around a guessed set of parameters with $\Gamma$, $\Delta E$ and $\Delta\mu$ of the order of $\kBT$, and for a strongly coupled tip, $\epsilon\approx1/2$. 
The obtained parameters give us information about how the optimal transmission probabilities result from the combination of the interference patterns and the properties of the scatterers. The resulting transmissions are plotted in Fig.~\ref{fig:opt_trans} for the QD and DQD cases for two different $(d-x,x)$ configurations chosen to have a maximal power. In both cases, we observe that the transmission probabilities between the conductor terminals and H are broad peaks with sharp borders. This is particularly evident in the DQD case, for which both ${\cal T}_{LH}(E)$ and ${\cal T}_{RH}(E)$ approach a boxcar function of width $\sim8\kBT$ and centered around $E-\mu\approx\Delta E/2\sim6\kBT$. 
Furthermore, as discussed in Sec.~\ref{sec:compose}, destructive interference almost completely cancels ${\cal T}_{LR}(E)$, confirming the detrimental contribution of elastic transport between the two conductor terminals. Clearly, electrons that are not absorbed by the tip do not contribute to the thermoelectric current. 

\begin{figure}[t]
    \centering
    \includegraphics[width=0.6\linewidth]{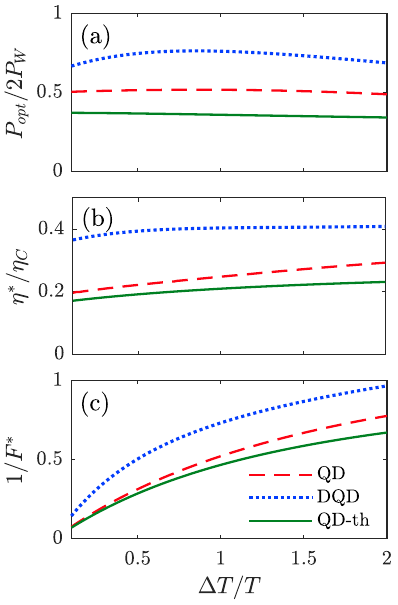}
    \caption{\label{fig:performance_comparison} Performance quantifiers: (a) $P_{opt}$, (b) $\eta^*$, and (c) $1/F^*$, as functions of the temperature difference for the QD (red dashed) and DQD (blue dotted) cases, compared with the QD-th (green full lines). The parameters are optimized for each system to extract maximum power at each $\Delta T$ and fixed $\mu=20\kBT$, with $x$ and $d$ being the same as in Fig.~\ref{fig:opt_trans}(a) and Fig.~\ref{fig:opt_trans}(b) for the QD and DQD, respectively.  }
\end{figure}

The chosen performance quantifiers for the different configurations (QD, DQD and QD-th) are shown in Fig.~\ref{fig:performance_comparison} by optimizing the parameters as $\Delta T$ is increased: $P_{opt}$, $\eta^*\equiv\eta(\Lambda_{opt})$ and $1/F^*$, with $F^*\equiv F(\Lambda_{opt})$. It shows that indeed the DQD setting performs better in terms of a larger power and efficiency, as well as a less noisy output current. The effect of the coherent coupling to H via the tip also results in being advantageous in the comparison of the QD and the QD-th setups. This is a clear manifestation of the improvement of considering the interference patterns that modulate the shape of the scattering resonances. 
The fact that the DQD is closer to a boxcar function than the QD is not totally surprising, as these transmissions have been proposed to be the asymptotic limit of arrays of QDS~\cite{whitney_most_2014,ehrlich_broadband_2021}. In our case, the Fabry-P\'erot like resonance between the DQD and the tip plays the role of an additional QD. An advantage of our setup is that, while one can only have control of the system parameters ($\varepsilon_\alpha$, $\Gamma$ and $\lambda$) up to some extent in a real experiment, sample imperfections can be compensated with an appropriate tuning of the tip position. It is expected that adding more QDs will help the optimization. However, this would come at the expense of increasing the system complexity and decreasing the degree of control.

As expected, the maximal power in general increases with $\Delta T$. Note that Fig.~\ref{fig:performance_comparison}(a) normalizes it with the power bound, cf. Eq.~\eqref{eq: power_bound}, which increases quadratically. For small temperature difference, $P_{opt}$ increases as $(\Delta T)^\gamma$, with $\gamma>2$ in the DQD, getting closest to $2P_W$ around $\Delta T\approx T$. In the other two cases, this ratio does not change appreciably, indicating $\gamma\approx2$, see Fig.~\ref{fig:performance_comparison}(a). Differently, the efficiency increases at the same rate as the Carnot efficiency for the DQD, while the QD and QD-th cases both increase their efficiency with $\Delta T$, see Fig.~\ref{fig:performance_comparison}(b). 

\begin{figure}[t!]
    \centering
    \includegraphics[width=\linewidth]{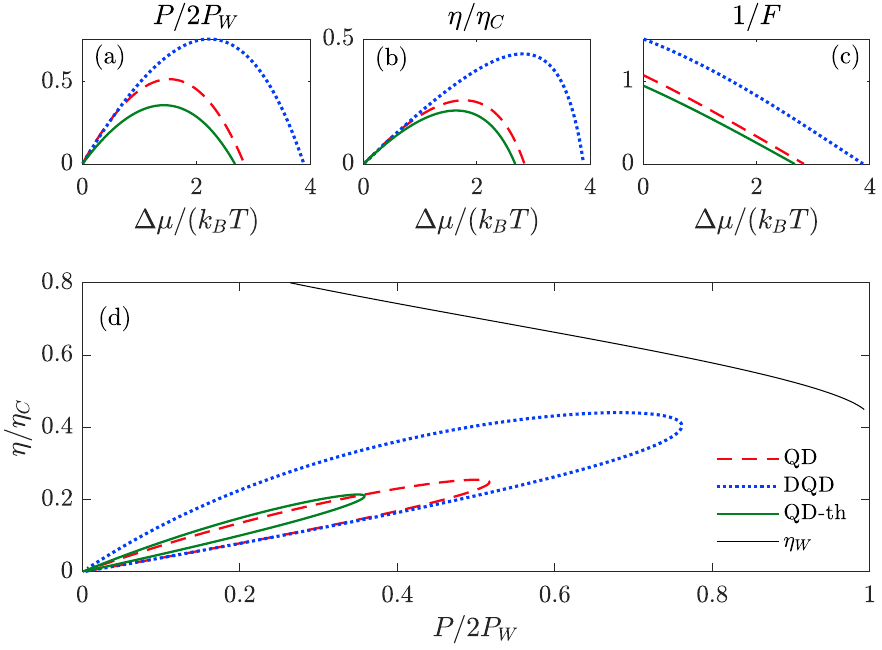}
    \caption{\label{fig:performance_cuts} Performance quantifiers: (a) $P$, (b) $\eta$, and (c) $1/F$, as functions of $\Delta\mu$ for the QD (red dashed) and DQD (blue dotted) cases, compared with the QD-th one (green full lines). The parameters are those that maximize the extracted power at $\Delta T/T=1$ and $\mu=20k_BT$ in each case: those of Fig. \ref{fig:opt_trans} for the QD and DQD cases, and $\Gamma=1.6\kBT$ and $\Delta E = 9\kBT$ for the QD-th. (d) Lasso diagrams relating $P$ and $\eta$ as $\Delta\mu$ is varied between 0 and the corresponding stopping voltage. The black curve indicates the calculated bound for efficiency at given power output, $\eta_W$.}
\end{figure}

We then chose the temperature difference $\Delta T=T$, and compare the performance of the different models (properly optimized) as a function of the applied bias. The results are shown in Figs.~\ref{fig:performance_cuts}(a)-(c). Again, the DQD case outperforms the other cases for all quantifiers. However, power and efficiency are very similar for QD and DQD at low voltages. This is not the case for the inverse Fano factor, for which QD and QD-th give lower values. As anticipated in Sec.~\ref{sec:response}, the Fano factor changes from sub to super-Poissonian with $\Delta\mu$, see Fig.~\ref{fig:performance_cuts}(c). Let us note that the DQD configuration seems to be beneficial for the increase of $1/F$ when compared with sharp step transmissions (with e.g., a quantum point contact) in two-terminal thermoelectrics~\cite{kheradsoud:2019}.  It is also useful to plot the power and efficiency relation as voltage is tuned between $\Delta\mu=0$ and the stopping voltage, see Fig.~\ref{fig:performance_cuts}(d). In all cases, the resulting elongated lasso diagrams involve that the maximum power and maximum efficiency points occur for not very different voltages. Remarkably, the efficiency of the DQD case reaches high values $\eta>0.4\eta_C$, not far from the maximal efficiency bound, $\eta_W$, (calculated following Ref. \cite{whitney_finding_2015}) for the same voltage (around $0.6\eta_C$). 

\begin{figure*}[t]
    \includegraphics[width=0.48\linewidth]{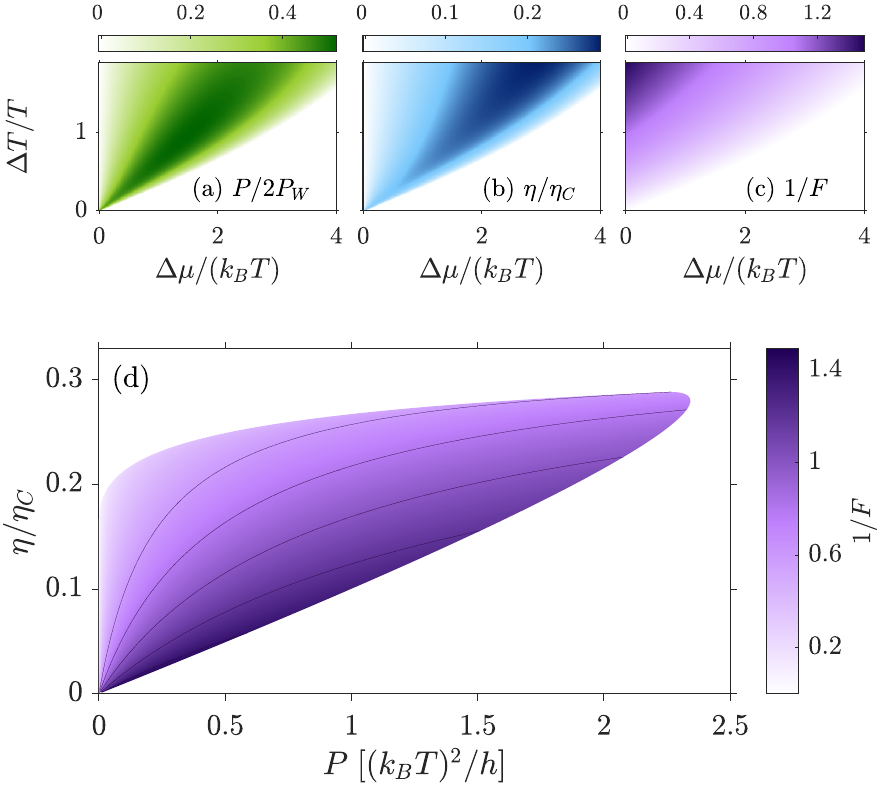}\qquad
    \includegraphics[width=0.48\linewidth]{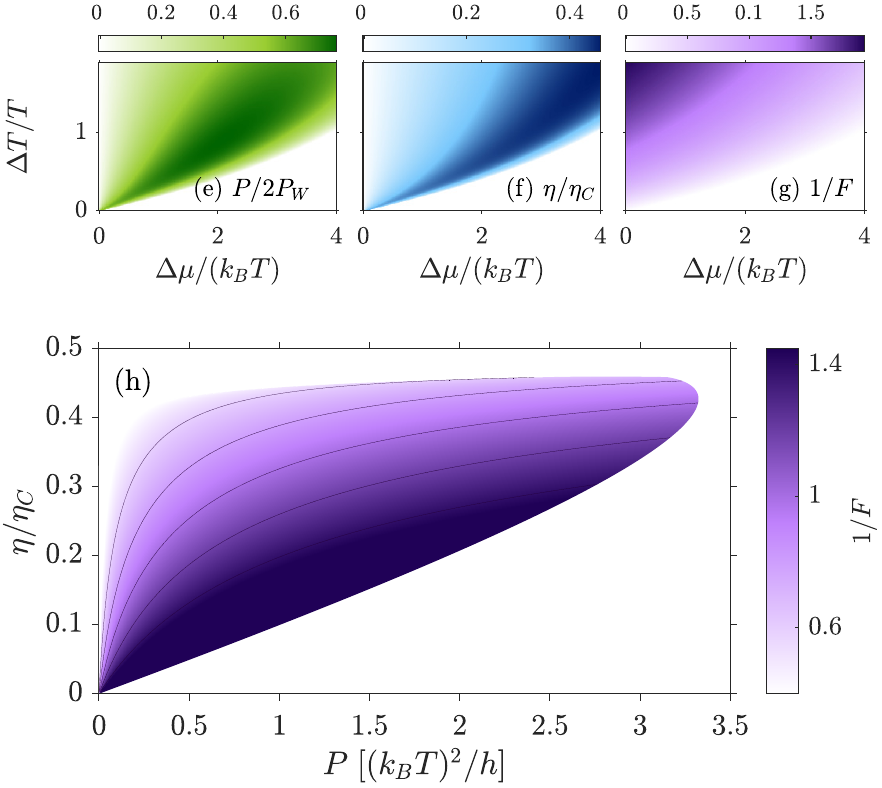}
   \caption{\label{fig:performance_QD} Performance quantifiers for (a)-(d) the QD and (e)-(h) the DQD configurations as functions of applied bias and temperature difference: (a), (e) Extracted power normalized by the power bound $P_W$; (b), (f) efficiency and (c), (g) inverse Fano factor. The parameters are those of Fig. \ref{fig:opt_trans}(a) for the QD case and of Fig. \ref{fig:opt_trans}(b) for the QD, chosen in both cases to maximize the power for fixed $\Delta T/T{=}1$ and $\mu{=}20\kBT$. (d), (h) Trade-off relation among $P$, $\eta$ and $1/F$ for different values of $\Delta \mu$ and $\Delta T$ (similar to the lasso diagrams of Fig. \ref{fig:performance_cuts}). Lines are shown along which $1/F$ is constant. }
\end{figure*}

Figure~\ref{fig:performance_QD} gives a more complete description of how the QD and DQD configurations perform as the system is brought out of equilibrium by increasing $\Delta\mu$ and $\Delta T$. The regions in which power and efficiency are maximal roughly coincide, though the maxima do not occur for the same configurations, see Figs.~\ref{fig:performance_QD}(a), ~\ref{fig:performance_QD}(b), ~\ref{fig:performance_QD}(e), and ~\ref{fig:performance_QD}(f). In this sense, it does not seem complicated to find a reasonable compromise to get a powerful and efficient heat engine. Reducing the output noise at the same time is more challenging, as $1/F$ increases in the region with small $\Delta\mu$ and high $\Delta T$ where both $P$ and $\eta$ are strongly suppressed, see Figs.~\ref{fig:performance_QD}(c) and \ref{fig:performance_QD}(g). Panels in Figs.~\ref{fig:performance_QD}(d) and \ref{fig:performance_QD}(h) may be helpful for that task. They can be understood as plotting $1/F$ in a compilation of all the lasso diagrams [like those in Fig.~\ref{fig:performance_cuts}(d)] for all $\Delta T$ giving a thermoelectric response. They clearly show that the signal-to-noise ratio is actually larger in the region where $P$ and $\eta$ are smallest. One can also see that for a given efficiency, $1/F$ increases with $P$. On the contrary, for a given power, $1/F$ decreases with $\eta$. Hence, to find a regular energy harvester, it seems more convenient to operate the system at higher powers even if efficiency is slightly compromised. Note that the DQD allows one to do so with a smaller decrease in the engine efficiency, as well as with an overall larger $P$ and $\eta$ approaching $\eta_C/2$.

\section{Conclusions}
\label{sec:conclusions}

We have investigated the role of internal coherence in the propagation of electrons through mesoscopic three-terminal energy harvesters. For this, we couple an isothermal two-terminal conductor to a heat bath via a coherent beam splitter, which allows for interference in the conductor-bath coupling. Experimentally, this can consist of a one-dimensional semiconductor quantum wire, with the QDs formed by conveniently stopping the growth process to include tunneling barriers~\cite{barker_individually_2019,dorsch:2020}. Side plunger gates can be then added to tune the QD energy levels. The coupling can be due to the tip of a scanning probe~\cite{chen_electron_2012,zolotavin:2017,fast:2020,gachter_spatially_2020}, hence introducing a mechanism to control the interference pattern. One-dimensional channels with QDs can also be patterned in two-dimensional materials~\cite{harzheim:2018} or two dimensional electron gases, see e.g., Refs.~\cite{talyanskii_single_1997,hermelin_electrons_2011}. The latter are unfortunately less accessible by a movable tip, though a hot third terminal can be connected by a quantum point contact at a fixed position of the wire~\cite{molenkamp:1990,jaliel:2019}. 

We find that the combination of resonant tunneling scatterers between the tip and the conductor terminal and modulated internal interferences considerably increases both the generated power and efficiency with respect to related configurations involving only one of these effects~\cite{extrinsic,jordan:2013}. This is due to the particular spectral dependence of the resulting transmission probabilities between the three terminals. Our results show that improving the (nonlocal) thermoelectric response requires the suppression of elastic transport along the conductor, which results from destructive interference for certain positions of the tip. On top of that, optimal configurations are found where the coupling between the conductor terminals and the tip occurs via broadband filters, in line with related proposals~\cite{whitney_most_2014}. Configurations with DQD scatterers give transmission probabilities which approach sharp boxcar functions, optimizing the heat engine performance in terms of the generated power, the efficiency and the output signal to noise ratio.  

As expected, one needs to find a compromise between producing the highest power and doing it at the highest efficiency. However, we observe the detriment in either case is not big. Though the signal-to-noise ratio is mostly enhanced in the regions with low power and low efficiency, it is found to improve for configurations with increasing power for a fixed efficiency. On the contrary, for a given power production, increasing the efficiency comes with the cost of noisier currents. 

Tunable internal interferences hence introduce a valuable way to control the properties of quantum heat engines via the coherent control of transport spectral properties. Here we discussed energy harvesting, but it will also affect absorption refrigerators~\cite{manikandan_autonomous_2020} as well, where heat autocorrelations will be important. Additional control may be achieved in interferometers that combine kinetic phases with magnetic field effects~\cite{haack:2019} or additional degrees of freedom like e.g., spin~\cite{puttock_local_2022}. 

Concerning the noise, auto- and crosscorrelations are sensitive to Coulomb interaction effects that are not treated at the mean-field level.  In particular, super-Poissonian autocorrelations are expected in the Coulomb blockade regime~\cite{cottet_positive_2004,sanchez:2013}  which may arise in the weakly coupled short wire limit. This effect is naively expected to be detrimental to the performance of the engine for increasing the noise at the same time reducing the current.

We have assumed here a phenomelogical approach based on simple but physically relevant scattering matrices describing resonant tunneling scatterers in a one-dimensional elastic conductor, considering that otherwise energy-dependent scattering only emerges due to phase coherence. Possible additional contributions due to e.g., the effect of disorder in the wire potential\footnote{See Refs.~\cite{bosisio_using_2015,muttalib_nonlinear_2015} for related works discussing diffusive thermoelectricity in disordered nanowires.} or of more involved energy dependence in the transmission coefficients simply add a level of complexity to the optimization problem but do not compromise our general conclusions. Extensions of our paper including the effect of dephasing~\cite{extrinsic} or of inelastic scattering due to the (typically weak) interaction with a thermal bath~\cite{pekola_colloquium_2021,wang_inelastic_2022} (phononic or electromagnetic environments) are interesting questions to be addressed in future works. 

\acknowledgements
We thank Matteo Acciai for useful comments on the paper, and acknowledge funding from the Ram\'on y Cajal Program No. RYC-2016-20778, and the Spanish Ministerio de Ciencia e Innovaci\'on via Grants No. PID2019-110125GB-I00 and No. PID2022-142911NB-I00, and through the Mar\'{i}a de Maeztu Programme for Units of Excellence in R{\&}D No. CEX2018-000805-M.

\appendix

\section{Double quantum dot scattering matrix}
\label{sec:dqdscatt}

The transmission and reflection amplitudes for a DQD are obtained via noninteracting single-particle Green's functions \cite{Datta1995,Ryndyk2016}. We model the scattering region defined by the coupled QDs by a two-site Hamiltonian with same level energy, $\varepsilon$, and coherent tunneling $\lambda$:
\begin{equation}
\label{eq: DQD hamiltonian}
   \hat{H}_{DQD}= \begin{pmatrix}
    \varepsilon & \lambda \\
    \lambda & \varepsilon
\end{pmatrix}.
\end{equation}
Coupling to left ($L'$) and right ($R'$) one-dimensional semi-infinite leads enters in the calculation of the retarded Green's function via self-energies $\hat{\Sigma}^r_{L'}$, $\hat{\Sigma}^r_{R'}$, 
\begin{equation}
\label{eq: GF_def}
    \hat{G}^r_S(E)=[E\hat{I}-\hat{H}_{DQD}-\hat{\Sigma}_{L'}^r-\hat{\Sigma}_{R'}^r]^{-1},
\end{equation}
with the identity matrix $\hat{I}$. In the wide-band limit and for symmetric coupling to the leads the self-energy matrices read:
\begin{equation}
\label{eq: self_energies}
    \hat{\Sigma}^r_{L'}= \begin{pmatrix}
                        -i\Gamma/2 & 0 \\
                        0 & 0
                      \end{pmatrix}\quad\text{and}\quad
    \hat{\Sigma}^r_{R'}= \begin{pmatrix}
                        0 & 0 \\
                        0 & -i\Gamma/2
                      \end{pmatrix},                      
\end{equation}
where $\Gamma = \hbar v$, $v$ being the velocity of the electrons in the leads. Self-energies beyond the wide-band limit would introduce an additional modulation to the scattering coefficients that can be compensated by the position of the tip or the QD levels~\cite{vanderWiel_electron_2002}. Substituting Eqs. \eqref{eq: DQD hamiltonian} and \eqref{eq: self_energies} in Eq. \eqref{eq: GF_def}, we obtain the desired Green's function of the system:
\begin{align}
\label{eq: green_function}
    \hat{G}^r_S(E)= \frac{1}{(E{-}\varepsilon {+} i\Gamma/2)^2{-}\lambda^2}
        \begin{pmatrix}
            E{-}\varepsilon {+}i\Gamma/2& \lambda \\
            \lambda & E{-}\varepsilon {+}i\Gamma/2
        \end{pmatrix}        .              
\end{align}
The scattering matrix can now be calculated from the retarded Green's function using the Fisher-Lee relation \cite{Datta1995,Fisher1981Jun}:
\begin{equation}
    {\cal S}(E)=\hat{I}-i\hbar v \hat{G}^r_S(E),
\end{equation}
the elements of which are the transmission and reflection amplitudes of Eq.~\eqref{eq:tau_dqd}.




\bibliography{biblio.bib}

\end{document}